\documentclass[12pt]{iopart}
\usepackage{graphicx}
\def\braket#1{\langle #1 \rangle}
\begin{document}

\title{Excitation spectrum and momentum distribution of
the ionic Bose-Hubbard model}

\author{Hiroki Nishizawa}

\address{Department of Applied Physics, The University of Tokyo, Hongo, Tokyo 113-8656, Japan}
\ead{hiroki-nishizawa752@g.ecc.u-tokyo.ac.jp}
\vspace{10pt}

\begin{abstract}
We investigate the excitation spectrum and momentum distribution of the ionic Bose-Hubbard model by the standard basis operator method. We derive Green's functions in the random phase approximation in Mott insulator, superfluid, charge density wave, and supersolid phases. The excitation spectrum has gapped modes and gapless Goldstone modes in the superfluid and supersolid phases. We show that the momentum distribution has a peak at the zone corner in the supersolid phase and the charge density wave phase close to the phase boundary. In addition, we demonstrate that the momentum distribution can be explained by the excitation spectrum and spectral weights of hole excitation modes.
\end{abstract}
\vspace{2pc}
\noindent{\it Keywords}: Bose-Hubbard, Green's function, supersolid, momentum distribution 

\section{Introduction}
The Bose-Hubbard model (BHM)~\cite{fisher1989boson, jaksch1998cold} describes bosonic atoms in optical lattices. Since parameters of optical lattices can be tuned flexibly, the BHM is ideal for the investigation of many-body effects. This model undergoes the phase transition from a superfluid (SF) phase to a Mott insulator (MI) phase at the critical ratio of the onsite repulsion to the hopping amplitude~\cite{freericks1994phase,van2001quantum,capogrosso2007phase}. These phases are characterized by the excitation spectrum. In the MI phase, all excitation modes are gapped. In contrast, in the SF phase, the spectrum consists of gapped and gapless Goldstone modes~\cite{sheshadri1993superfluid,ohashi2006itinerant,menotti2008spectral,huber2007dynamical,krutitsky2011excitation,dutta2012projection,di2018particle,caleffi2020quantum,sengupta2005mott}. The momentum distribution, which can be measured by time-of-flight spectroscopy, is also an important indicator of the SF-MI phase transition. In the SF phase, the momentum distribution has a sharp peak at zero momentum, which indicates the existence of Bose-Einstein condensation and superfluidity~\cite{menotti2008spectral,sengupta2005mott,kato2008sharp,freericks2009strong,knap2010spectral}. The excitation spectrum and momentum distribution provide important information about the phase of the BHM.

In addition to the MI and SF phase, there are solid-ordered phases: the phases with a periodic density modulation such as a charge density wave (CDW) and supersolid (SS) phase. The properties of solid-ordered phases have been discussed in studies of the extended Bose-Hubbard model~\cite{batrouni1995supersolids,sengupta2005supersolids,kovrizhin2005density,yamamoto2009successive,iskin2011route,kimura2011gutzwiller,trefzger2011ultracold,rossini2012phase,ohgoe2012commensurate,dutta2015non}. Although there are a few studies on the excitation spectrum~\cite{kovrizhin2005density,gremaud2016excitation} and the momentum distribution~\cite{iskin2009momentum,ohgoe2012ground}, the nearest-neighbor repulsion makes it difficult to investigate generally these properties. Another way to form solid-ordered phases is to introduce the staggered potential.
The BHM with the staggered potential is called the ionic Bose Hubbard model (IBHM)~\cite{guo2009cold,sawhney2021influence}. 
The IBHM is suitable for investigating solid-ordered phases, but few studies have been carried out on the IBHM.
The excitation spectrum of the IBHM has been calculated by the spin-wave approximation in the hard-core limit~\cite{li2015hard}. However, only the half-filling case has been considered. Moreover, there is no theoretical study of the momentum distribution of the IBHM.

The purpose of this study is to investigate the excitation spectrum and momentum distribution of the IBHM on a square lattice by the standard basis operator (SBO) method. The SBO method has been applied to the BHM in both the  MI and SF phases~\cite{sheshadri1993superfluid,ohashi2006itinerant,menotti2008spectral,sajna2015ground}. We extend the SBO method to the IBHM on a bipartite square lattice, which has two sublattices due to a periodic density modulation. 
In Sec.~\ref{sec:mean}, we summarize the mean-field phase diagram of the IBHM. In Sec.~\ref{sec:green}, we derive equations for Green's functions by the SBO method in the random phase approximation. By solving the equations, we obtain the Green's functions in the MI, SF, CDW, and SS phases.  In Sec.~\ref{sec:ex}, we calculate the excitation spectrum that has gapped and gapless modes in the SF and SS phases.  In Sec.~\ref{sec:mo}, we show that the momentum distribution has a peak at the zone corner in the SS phase and the CDW phase close to the CDW-SS phase transition. In addition, we demonstrate that the momentum distribution can be understood by the excitation spectrum and spectral weights of hole excitations.  

\section{Mean-field phase diagram}
\label{sec:mean}

In this section, we summarize the phase diagram of the ionic Bose-Hubbard model
\begin{equation}
\label{ham}
H=-t\sum_{\braket{ij}}(b_{i}^{\dagger}b_{j}+\mathrm{H.c.})+\frac{U}{2}\sum_{i}n_{i}(n_{i}-1)-\mu\sum_{i} n_i-\Delta\sum_{i\in A}n_{i}+\Delta\sum_{i\in B}n_{i},
\end{equation}
where $b_{i}^{\dagger}$ and $b_{i}$ are the boson creation and annihilation operators at site $i$, $n_{i}=b_{i}^{\dagger}b_{i}$ is the number operator, $t$ is the hopping amplitude, $\braket{ij}$ denotes the summation over the nearest-neighbor sites, $U$ is the onsite repulsion, $\mu$ is the chemical potential, and $\Delta$ is the staggered potential on sublattices A and B. 

To study the phase diagram, we apply the mean-field approximation~\cite{yamamoto2009successive}. In this approximation, the hopping term is decoupled as $b_{i}^{\dagger}b_{j}\simeq b_{i} \braket{b_j} + \braket{b_i} b_j$. The mean-field Hamiltonian is
\begin{eqnarray}
&H_{\mathrm{MF}}=\sum_{i \in A}H_{A}+\sum_{i \in B}H_{B}, \\
H_{A}=-zt\phi_B&(b_{i}^{\dagger}+b_{i})+\frac{U}{2}n_{i}(n_{i}-1)-(\mu+\Delta) n_{i},  \\
H_{B}=-zt\phi_A&(b_{i}^{\dagger}+b_{i})+\frac{U}{2}n_{i}(n_{i}-1)-(\mu-\Delta) n_{i}.
\end{eqnarray}
where  $z$ is the lattice coordination number ($z=4$ for a square lattice), $\phi_A=\braket{b_i}_{i \in A}=\braket{b_{i}^{\dagger}}_{i \in A}$ and $\phi_B=\braket{b_i}_{i \in B}=\braket{b_{i}^{\dagger}}_{i \in B}$ are the expectation values of the annihilation operators on sublattices $A$ and $B$, respectively. Likewise, $n_A=\braket{n_{i}}_{i \in A}$ and $n_B=\braket{n_{i}}_{i \in B}$ are the expectation values of the number operators on sublattices $A$ and $B$, respectively. The expectation values are calculated numerically using the eigenstates of $H_{A}$ and $H_{B}$. 
\begin{figure*}
\includegraphics{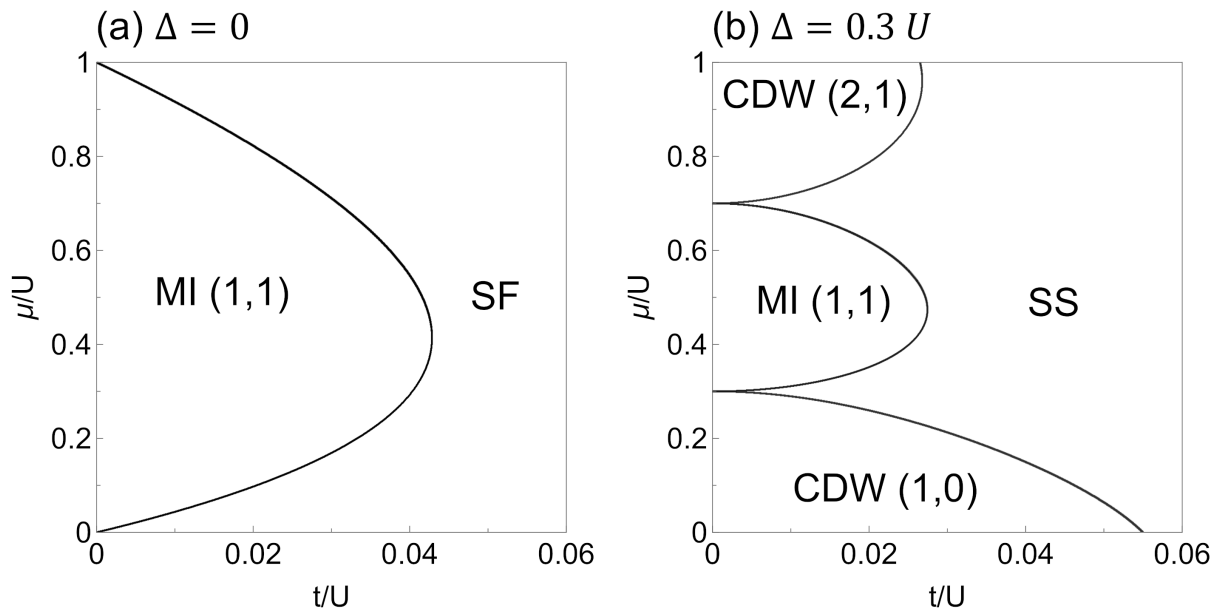}
\caption{\label{phase}Phase diagrams of the ionic Bose-Hubbard model on a square lattice ($z=4$) at zero temperature by the mean-field approximation  for (a) $\Delta=0$ and (b) $\Delta=0.3\,U$. The MI and CDW phases are labeled by $(n_A,n_B)$, which represents the expectation values of number operators on sublattices $A$ and $B$.}
\end{figure*}

The phase diagrams of the ionic Bose-Hubbard model on a square lattice are shown in \fref{phase}. 
The phase diagram for $\Delta=0$ agrees with that for the standard Bose-Hubbard model. There are MI phases ($\phi_A=\phi_B=0$, $n_A = n_B$) and the SF phase ($\phi_A=\phi_B>0$). For $\Delta=0.3\,U$,  CDW phases ($\phi_A=\phi_B=0$, $n_A \neq  n_B$) and the SS phase ($\phi_A\neq \phi_B$) exist.
In the atomic limit ($t=0$), the expectation value $n_A$ is given by the integer $n$ that minimizes the energy $E_A=Un(n-1)/2-(\mu+\Delta) n$. Thus,  $n_A=n$ for $U(n-1)-\Delta<\mu< Un-\Delta$. Likewise, $n_B=n$ for $U(n-1)+\Delta<\mu<Un+\Delta$. Due to the staggered potential $\Delta\neq0$, the ground state alternates between CDW and MI phases as $\mu$ increases.
When $t$ increases, MI and CDW regions shrink, and the SS phase appears.

\section{Green's function}
\label{sec:green}
To calculate the momentum dependence of the excitation energy and distribution function, we derive Green's functions by the standard basis operator method~\cite{sheshadri1993superfluid,ohashi2006itinerant,menotti2008spectral,iskin2009momentum,sajna2015ground,konabe2006laser,haley1972standard}. By considering fluctuations around the mean-field ground state, the Hamiltonian is rewritten as 
\begin{equation}
\label{hl}
H=H_{\mathrm{MF}}-t\sum_{\langle i,j \rangle}(\delta  b_{i}^{\dagger}\delta b_{j}+\mathrm{H.c.}),
\end{equation}
where the operators $\delta b_{i}^{\dagger}=b_{i}^{\dagger}-\phi_i$ and $\delta b_{i}=b_{i}-\phi_i$ represent the deviations from the mean-field Hamiltonian. By using the energy eigenstates $\vert i,\alpha\rangle$ of the mean-field Hamiltonian at site $i$ with eigenenergies $E^{i}_{\alpha}$, equation \eref{hl} is rewritten as
\begin{equation}
H=\sum_{i,\alpha}E^{i}_{\alpha}\hat{L}^{i}_{\alpha \alpha} -t\sum_{\langle i,j \rangle }\sum_{\alpha \alpha^{\prime} }\sum_{\beta \beta^{\prime}}(d^{i}_{\alpha \alpha^{\prime} }c^{j}_{\beta \beta^{\prime}}\delta \hat{L}^{i}_{\alpha \alpha^{\prime}}\delta \hat{L}^{j}_{\beta \beta^{\prime}}+\mathrm{H.c.}),
\end{equation}
where $\hat{L}^{i}_{\alpha \alpha^{\prime}}=\vert i,\alpha\rangle \langle i , \alpha^{\prime} \vert$ is the standard basis operator (SBO)~\cite{haley1972standard}, $\delta \hat{L}^{i}_{\alpha \alpha^{\prime}}=\hat{L}^{i}_{\alpha \alpha^{\prime}}-\braket{ \hat{L}^{i}_{\alpha \alpha^{\prime}} }$, $d^{i}_{\alpha \alpha^{\prime} }=\langle i,\alpha \vert b_{i}^{\dagger} \vert i, \alpha^{\prime} \rangle$, and $c^{j}_{\beta \beta^{\prime}}=\langle j,\beta \vert b_{j} \vert j, \beta^{\prime} \rangle$. In this formalism,  the retarded Green's function
$G^{ij}(t-t')=-i\Theta(t-t')\braket{[b_i(t),b_j^{\dagger}(t')]}$~\cite{zubarev1960double}
is written as
\begin{equation}
\label{ij}
G^{ij}(t-t')=\sum_{\alpha \alpha^{\prime} }\sum_{\beta \beta^{\prime}}c^{i}_{\alpha \alpha^{\prime} }d^{j}_{\beta \beta^{\prime}}G^{ij}_{\alpha \alpha^{\prime},\beta \beta^{\prime}}(t-t'),
\end{equation}
where
\begin{equation}
G^{ij}_{\alpha \alpha^{\prime},\beta \beta^{\prime}}(t-t')=-i\Theta(t-t')\braket{[\hat{L}^{i}_{\alpha \alpha^{\prime}}(t), \hat{L}^{j}_{\beta \beta^{\prime}}(t')]}.
\end{equation}
The equation for the Green's function in frequency space in the random phase approximation (RPA)~\cite{sheshadri1993superfluid,ohashi2006itinerant,menotti2008spectral,iskin2009momentum,sajna2015ground,konabe2006laser} is
\begin{equation}
\label{eq}
[\omega-E^{i}_{\alpha' \alpha}]G^{ij}_{\alpha \alpha',\beta \beta'}(\omega)=P^{i}_{\alpha \alpha'}\delta_{\alpha' \beta}\delta_{\alpha \beta'}\delta^{ij}-tP^{i}_{\alpha \alpha'}\sum_{l}\sum_{\gamma \gamma'}T^{il}_{\alpha \alpha' \gamma \gamma'}G^{lj}_{\gamma \gamma',\beta \beta'}(\omega),
\end{equation}
where
$E^{i}_{\alpha' \alpha}=E^{i}_{\alpha'}-E^{i}_{\alpha}$,
$T^{il}_{\alpha \alpha' \gamma \gamma'}=d^{i}_{\alpha \alpha'}c^{l}_{\gamma \gamma'}+c^{i}_{\alpha \alpha'}d^{l}_{\gamma \gamma'}$,
$P^{i}_{\alpha \alpha'}=P^{i}_{\alpha}-P^{i}_{\alpha'}$, and $P^{i}_{\alpha}=\braket{\hat{L}^{i}_{\alpha \alpha}}$ is the occupation probability. This quantity is equal to unity when $\alpha$ is the ground state, and it is zero when $\alpha$ is another state at zero temperature. By solving equation \eref{eq}, we obtain the Green's function $G_{\mathbf{k}}(\omega)$, which is expressed by the mean-field local Green's functions given by
\begin{equation}
\left(
\begin{array}{cc}
F^{i}_{cd} & F^{i}_{cc} \\
F^{i}_{dd} & F^{i}_{dc} \\
\end{array}
\right) =\sum_{\alpha \alpha'}
\frac{P^{i}_{\alpha \alpha'}}{\omega-E^{i}_{\alpha' \alpha}}
\left(
\begin{array}{cc}
c^{i}_{\alpha \alpha'}d^{i}_{\alpha' \alpha} & c^{i}_{\alpha \alpha'}c^{i}_{\alpha' \alpha} \\
d^{i}_{\alpha \alpha'}d^{i}_{\alpha' \alpha} & d^{i}_{\alpha \alpha'}c^{i}_{\alpha' \alpha} \\
\end{array} 
\right).
\end{equation}
We solve equation \eref{eq} and calculate the excitation spectrum and momentum distribution in the MI, SF, CDW, and SS phases. First, we consider Green's functions for the staggered potential $\Delta =0$. Then, we derive Green's functions for $\Delta\neq0$.

\subsection{MI ($\Delta=0$)}
For $\Delta=0$, sublattices $A$ and $B$ are equivalent. Thus, the Fourier transformation of the Green's function in equation \eref{ij} is defined as 
\begin{eqnarray}
G^{ij}(\omega)=&\frac{1}{N}\sum_{\mathbf{k}}G_{\mathbf{k}}(\omega)e^{-i\mathbf{k}\cdot(\mathbf{R_i}-\mathbf{R_j})}, \\
G_{\mathbf{k}}(\omega)=&\frac{1}{N}\sum_{ij}G^{ij}(\omega)e^{i\mathbf{k}\cdot(\mathbf{R_i}-\mathbf{R_j})},
\end{eqnarray}
where $N$ is the number of lattice sites. In momentum space, the equation for the Green's function is
\begin{equation}
\label{eq:mott}
G_{\mathbf{k}}(\omega)=F_{cd}+\epsilon_{\mathbf{k}}F_{cd}G_{\mathbf{k}}(\omega),
\end{equation}
where $\epsilon_{\mathbf{k}}=-2t\sum_{i=x,y}\cos(k_i)$, and the lattice constant is set to unity.
From equation \eref{eq:mott}, the Green's function is
\begin{equation}
\label{G}
G_{\mathbf{k}}(\omega)=F_{cd}/(1-\epsilon_{\mathbf{k}}F_{cd}).
\end{equation}
In the MI phase, $G_{\mathbf{k}}(\omega)$ is analytically obtained since $F_{cd}$ is given analytically. Considering that the ground state is the number state $|i,n_i\rangle$ with $E_{n_i}=U{n_i}({n_i}-1)/2-\mu {n_i}$ and $c^{i}_{n_i' n_i}=\sqrt{n_i}\delta_{n'_i,n_i-1}$, the mean-field Green's function is
\begin{equation}
\label{F}
F_{cd}=\frac{n+1}{\omega-E^{p}}-\frac{n}{\omega+E^{h}},
\end{equation}
where $E^{p}=E_{n+1}-E_{n}=Un-\mu$ and $E^h=E_{n-1}-E_{n}=-U(n-1)+\mu$ are particle and hole excitation energies, respectively. By substituting equation \eref{F} into equation \eref{G}, the Green's function is 
\begin{equation}
G_{\mathbf{k}}(\omega)=\frac{C^p(\mathbf{k})}{\omega-E^{p}(\mathbf{k})}-\frac{C^h(\mathbf{k})}{\omega+E^{h}(\mathbf{k})},
\end{equation}
where $E^{p}(\mathbf{k})=E^p-[U-E^U(\mathbf{k})-\epsilon_{\mathbf{k}}]/2$, $E^{h}(\mathbf{k})=E^h-[U-E^U(\mathbf{k})+\epsilon_{\mathbf{k}}]/2$, $E^U(\mathbf{k})=\sqrt{\epsilon_{\mathbf{k}}^2+2U(2n+1)\epsilon_{\mathbf{k}}+U^2}$,  $C^p(\mathbf{k})=[\epsilon_{\mathbf{k}}+(2n+1)U+E^U(\mathbf{k})]/[2E^U(\mathbf{k})]$, and $C^h(\mathbf{k})=[\epsilon_{\mathbf{k}}+(2n+1)U-E^U(\mathbf{k})]/[2E^U(\mathbf{k})]$. The spectral function is
\begin{eqnarray}
A(\mathbf{k},\omega)&=-\frac{1}{\pi}\mathrm{Im}G_{\mathbf{k}}(\omega+i0^{+}) \nonumber \\
&=C^p(\mathbf{k})\delta[\omega-E^{p}(\mathbf{k})]-C^h(\mathbf{k})\delta[\omega+E^{h}(\mathbf{k})],
\end{eqnarray}
where $C^p(\mathbf{k})$ is the spectral weight of the particle excitation, and $C^h(\mathbf{k})$ is that of the hole excitation.
The momentum distribution is
\begin{equation}
n(\mathbf{k})=-\int_{-\infty}^{0}A(\mathbf{k},\omega) d\omega=C^h(\mathbf{k}).
\end{equation}

\subsection{SF}
In the SF phase ($\phi_A=\phi_B\neq0$, $n_A = n_B$), it is necessary to introduce the anomalous Green's function~\cite{sajna2015ground} defined by 
\begin{equation}
H^{ij}(t-t')=-i\Theta(t-t')\braket{[b^{\dagger}_i(t),b_j^{\dagger}(t')]}
=\sum_{\alpha \alpha^{\prime} }\sum_{\beta \beta^{\prime}}d^{i}_{\alpha \alpha^{\prime} }d^{j}_{\beta \beta^{\prime}}G^{ij}_{\alpha \alpha^{\prime},\beta \beta^{\prime}}(t-t').
\end{equation}
The equations for the Green's functions are
\begin{eqnarray}
&G_{\mathbf{k}}(\omega)=F_{cd}+\epsilon_{\mathbf{k}}F_{cc}H_{\mathbf{k}}(\omega)+\epsilon_{\mathbf{k}}F_{cd}G_{\mathbf{k}}(\omega), \\
&H_{\mathbf{k}}(\omega)=F_{dd}+\epsilon_{\mathbf{k}}F_{dc}H_{\mathbf{k}}(\omega)+\epsilon_{\mathbf{k}}F_{dd}G_{\mathbf{k}}(\omega).
\end{eqnarray}
From these equations,
\begin{eqnarray}
&G_{\mathbf{k}}(\omega)=(F_{cd}-\epsilon_{\mathbf{k}}F_{cd}F_{dc}+\epsilon_{\mathbf{k}}F_{cc}F_{dd})/D_{\mathbf{k}}, \label{GSF}\\
&H_{\mathbf{k}}(\omega)=F_{dd}/D_{\mathbf{k}},
\end{eqnarray}
where 
\begin{equation}
D_{\mathbf{k}}=1-\epsilon_{\mathbf{k}}F_{cd}-\epsilon_{\mathbf{k}}F_{dc}+\epsilon^2_{\mathbf{k}}F_{cd}F_{dc}-\epsilon^2_{\mathbf{k}}F_{cc}F_{dd}.
\end{equation}
\Eref{GSF} is the same as
$G_{\mathbf{k}}(\omega)=\Pi_{\mathbf{k}}(\omega)/[1-\epsilon_{\mathbf{k}} \Pi_{\mathbf{k}}(\omega)]$, where $\Pi_{\mathbf{k}}(\omega)=F_{cd}+\epsilon_{\mathbf{k}}F_{cc}F_{dd}/(1-\epsilon_{\mathbf{k}} F_{dc})$~\cite{ohashi2006itinerant,menotti2008spectral,sajna2015ground}.
In contrast to the MI phase, the mean-field Green's functions can not be obtained analytically, so the Green's function is obtained numerically. To obtain the excitation energies and spectral weights, we factor the numerator and denominator of the Green's function and then cancel out common terms numerically. Consequently, the Green's function is decomposed as
\begin{equation}
G_{\mathbf{k}}(\omega)=\sum_{\alpha} \frac{C^p_\alpha(\mathbf{k})}{\omega-E^p_\alpha(\mathbf{k})}-\sum_{\alpha} \frac{C^h_\alpha(\mathbf{k})}{\omega+E^h_\alpha(\mathbf{k})}.
\end{equation}
The number of the poles of the Green's function is the same as that of the mean-field Green's functions. When the mean-field Hamiltonian is diagonalized in the occupation number basis $\{|0\rangle, |1\rangle, \ldots, |n_{\mathrm{max}}\rangle \}$, the number of the poles is $2 n_{\mathrm{max}}$ at zero temperature.
The momentum distribution is 
\begin{equation}
n(\mathbf{k})=\sum_{\alpha}C^h_\alpha(\mathbf{k})+N \phi^2 \delta_{\mathbf{k},\Gamma},
\end{equation}
where $\phi^2=\phi_A^2=\phi_B^2$ is the condensate fraction, $N \phi^2$ is the contribution of the condensate, and $\Gamma=(0,0)$ is the zone center, where the excitation energy is zero.

\subsection{CDW and MI ($\Delta\neq0$)}
For $\Delta\neq0$, sublattices $A$ and $B$ are inequivalent and each consists of $N/2$ lattice sites. Therefore, the Fourier transformation of Green's functions for sublattices is defined as 
\begin{eqnarray}
G^{i_mj_n}(\omega)=&\frac{2}{N}\sum_{\mathbf{k}}G^{mn}_{\mathbf{k}}(\omega)e^{-i\mathbf{k}\cdot(\mathbf{R_{i_m}}-\mathbf{R_{j_n}})}, \\
G^{mn}_{\mathbf{k}}(\omega)=&\frac{2}{N}\sum_{i_mj_n}G^{i_mj_n}(\omega)e^{i\mathbf{k}\cdot(\mathbf{R_{i_m}}-\mathbf{R_{j_n}})},
\end{eqnarray}
where indices $m$ and $n$ label sublattices $A$ and $B$~\cite{frobrich2006many}. The equations for the Green's functions in momentum space are
\begin{eqnarray}
&G^{AA}_{\mathbf{k}}(\omega)=F^{A}_{cd}+\epsilon_{\mathbf{k}}F^{A}_{cd}G^{BA}_{\mathbf{k}}(\omega), \\
&G^{BA}_{\mathbf{k}}(\omega)=\epsilon_{\mathbf{k}}F^{B}_{cd}G^{AA}_{\mathbf{k}}(\omega), \\
&G^{BB}_{\mathbf{k}}(\omega)=F^{B}_{cd}+\epsilon_{\mathbf{k}}F^{B}_{cd}G^{AB}_{\mathbf{k}}(\omega), \\
&G^{AB}_{\mathbf{k}}(\omega)=\epsilon_{\mathbf{k}}F^{A}_{cd}G^{BB}_{\mathbf{k}}(\omega).
\end{eqnarray}
From these equations, the Green's functions are
\begin{eqnarray}
&G^{AA}_{\mathbf{k}}(\omega)=F^{A}_{cd}/D_{\mathbf{k}}, \\
&G^{BA}_{\mathbf{k}}(\omega)=G^{AB}_{\mathbf{k}}(\omega)=\epsilon_{\mathbf{k}}F^{A}_{cd}F^{B}_{cd}/D_{\mathbf{k}}, \\
&G^{BB}_{\mathbf{k}}(\omega)=F^{B}_{cd}/D_{\mathbf{k}},
\end{eqnarray}
where $D_{\mathbf{k}}=1-\epsilon_{\mathbf{k}}^2F^{A}_{cd}F^{B}_{cd}$.
These Green's functions can be analytically expressed since $F^{A}_{cd}$ and $F^{B}_{cd}$ are given by
\begin{eqnarray}
F^{A}_{cd}=\frac{n_A+1}{\omega-E^p_A}-\frac{n_A}{\omega+E^h_A}, \\
F^{B}_{cd}=\frac{n_B+1}{\omega-E^p_B}-\frac{n_B}{\omega+E^h_B},
\end{eqnarray}
where $E^p_A=Un_A-(\mu+\Delta)$, $E^h_A=-U(n_A-1)+(\mu+\Delta)$, $E^p_B=Un_B-(\mu-\Delta)$, and $E^h_B=-U(n_B-1)+(\mu-\Delta)$.
The Green's function is
\begin{eqnarray}
G_{\mathbf{k}}(\omega)&=\frac{1}{N}\sum_{ij}G^{ij}(\omega)e^{i\mathbf{k}\cdot(\mathbf{R_i}-\mathbf{R_j})}=\frac{1}{2}\sum_{mn}G^{mn}_{\mathbf{k}}(\omega) \nonumber \\
&=\frac{1}{2}[G^{AA}_{\mathbf{k}}(\omega)+G^{AB}_{\mathbf{k}}(\omega)+G^{BA}_{\mathbf{k}}(\omega)+G^{BB}_{\mathbf{k}}(\omega)].
\end{eqnarray}
The excitation energies are the poles of the Green's function, which are the solutions of $D_{\mathbf{k}}=1-\epsilon_{\mathbf{k}}^2F^{A}_{cd}F^{B}_{cd}=0$. This is a quartic equation in $\omega$, so the poles are analytically obtained by Ferrari's formula. Thus, the Green's function is written as
\begin{equation}
G_{\mathbf{k}}(\omega)=\frac{C_A^p(\mathbf{k})}{\omega-E_A^{p}(\mathbf{k})}+\frac{C_B^p(\mathbf{k})}{\omega-E_B^{p}(\mathbf{k})} -\frac{C_A^h(\mathbf{k})}{\omega+E_A^{h}(\mathbf{k})}-\frac{C_B^h(\mathbf{k})}{\omega+E_B^{h}(\mathbf{k})},
\end{equation}
where the excitation energies and spectral weights can be analytically expressed, but the expressions are too long to write here. The momentum distribution is
\begin{equation}
n(\mathbf{k})=-\int_{-\infty}^{0}A(\mathbf{k},\omega) d\omega=C_A^h(\mathbf{k})+C_B^h(\mathbf{k}).
\end{equation}

\subsection{SS}
We derive Green's functions in the SS phase ($\phi_A\neq\phi_B$, $n_A \neq n_B$). The equations for the Green's functions are
\begin{eqnarray}
G^{BB}_{\mathbf{k}}(\omega)=&F^{B}_{cd}+\epsilon_{\mathbf{k}}F^{B}_{cd}G^{AB}_{\mathbf{k}}(\omega)+\epsilon_{\mathbf{k}}F^{B}_{cc}H^{AB}_{\mathbf{k}}(\omega), \label{GBB}\\
G^{AB}_{\mathbf{k}}(\omega)=&\epsilon_{\mathbf{k}}F^{A}_{cd}G^{BB}_{\mathbf{k}}(\omega)+\epsilon_{\mathbf{k}}F^{A}_{cc}H^{BB}_{\mathbf{k}}(\omega), \label{GAB}\\
H^{BB}_{\mathbf{k}}(\omega)=&F^{B}_{dd}+\epsilon_{\mathbf{k}}F^{B}_{dc}H^{AB}_{\mathbf{k}}(\omega)+\epsilon_{\mathbf{k}}F^{B}_{dd}G^{AB}_{\mathbf{k}}(\omega), \label{HBB} \\
H^{AB}_{\mathbf{k}}(\omega)=&\epsilon_{\mathbf{k}}F^{A}_{dc}H^{BB}_{\mathbf{k}}(\omega)+\epsilon_{\mathbf{k}}F^{A}_{dd}G^{BB}_{\mathbf{k}}(\omega), \label{HAB}\\
G^{AA}_{\mathbf{k}}(\omega)=&F^{A}_{cd}+\epsilon_{\mathbf{k}}F^{A}_{cd}G^{BA}_{\mathbf{k}}(\omega)+\epsilon_{\mathbf{k}}F^{A}_{cc}H^{BA}_{\mathbf{k}}(\omega), \\
G^{BA}_{\mathbf{k}}(\omega)=&\epsilon_{\mathbf{k}}F^{B}_{cd}G^{AA}_{\mathbf{k}}(\omega)+\epsilon_{\mathbf{k}}F^{B}_{cc}H^{AA}_{\mathbf{k}}(\omega), \\
H^{AA}_{\mathbf{k}}(\omega)=&F^{A}_{dd}+\epsilon_{\mathbf{k}}F^{A}_{dc}H^{BA}_{\mathbf{k}}(\omega)+\epsilon_{\mathbf{k}}F^{A}_{dd}G^{BA}_{\mathbf{k}}(\omega), \\
H^{BA}_{\mathbf{k}}(\omega)=&\epsilon_{\mathbf{k}}F^{B}_{dc}H^{AA}_{\mathbf{k}}(\omega)+\epsilon_{\mathbf{k}}F^{B}_{dd}G^{AA}_{\mathbf{k}}(\omega).
\end{eqnarray}
From equations (\ref{GBB})--(\ref{HAB}), the Green's functions are 
\begin{eqnarray}
G^{BB}_{\mathbf{k}}(\omega)=&(F^{B}_{cd}-\epsilon_{\mathbf{k}}^2F^{A}_{dc} F^{B}_{cd}  F^{B}_{dc}+\epsilon_{\mathbf{k}}^2F^{A}_{dc}F^{B}_{cc}F^{B}_{dd})/D_{\mathbf{k}}, \\ 
G^{AB}_{\mathbf{k}}(\omega)=&\epsilon_{\mathbf{k}}(F^{A}_{cd}F^{B}_{cd}+F^{A}_{cc}F^{B}_{dd}-\epsilon_{\mathbf{k}}^2F^{A}_{cd}F^{A}_{dc}F^{B}_{cd}F^{B}_{dc} \nonumber \\
&-\epsilon_{\mathbf{k}}^2F^{A}_{cc}F^{A}_{dd}F^{B}_{cc}F^{B}_{dd}+\epsilon_{\mathbf{k}}^2F^{A}_{cc}F^{A}_{dd}F^{B}_{cd}F^{B}_{dc}\nonumber \\
&+\epsilon_{\mathbf{k}}^2F^{A}_{cd}F^{A}_{dc}F^{B}_{cc}F^{B}_{dd})/D_{\mathbf{k}}, \\ 
H^{BB}_{\mathbf{k}}(\omega)=&(F^{B}_{dd}+\epsilon_{\mathbf{k}}^2F^{A}_{dd}F^{B}_{cd}F^{B}_{dc}-\epsilon_{\mathbf{k}}^2F^{A}_{dd}F^{B}_{cc}F^{B}_{dd})/D_{\mathbf{k}}, \\
H^{AB}_{\mathbf{k}}(\omega)=&\epsilon_{\mathbf{k}}(F^{A}_{dd}F^{B}_{cd}+F^{A}_{dc}F^{B}_{dd})/D_{\mathbf{k}},
\end{eqnarray} where
\begin{eqnarray}
D_{\mathbf{k}}=&1-\epsilon_{\mathbf{k}}^2 F^{A}_{cd} F^{B}_{cd}  -\epsilon_{\mathbf{k}}^2 F^{A}_{dc} F^{B}_{dc}  - \epsilon_{\mathbf{k}}^2 F^{A}_{dd} F^{B}_{cc} - \epsilon_{\mathbf{k}}^2 F^{A}_{cc} F^{B}_{dd} \nonumber \\ 
&+ \epsilon_{\mathbf{k}}^4 F^{A}_{cd} F^{A}_{dc} F^{B}_{cd} F^{B}_{dc} + \epsilon_{\mathbf{k}}^4 F^{A}_{cc} F^{A}_{dd} F^{B}_{cc} F^{B}_{dd} \nonumber \\
&- \epsilon_{\mathbf{k}}^4 F^{A}_{cc} F^{A}_{dd} F^{B}_{cd} F^{B}_{dc}  - \epsilon_{\mathbf{k}}^4 F^{A}_{cd} F^{A}_{dc} F^{B}_{cc} F^{B}_{dd}.
\end{eqnarray}
The Green's functions $G^{AA}_{\mathbf{k}}(\omega)$, $G^{BA}_{\mathbf{k}}(\omega)$, $H^{AA}_{\mathbf{k}}(\omega)$, and $H^{BA}_{\mathbf{k}}(\omega)$ are derived from $G^{BB}_{\mathbf{k}}(\omega)$, $G^{AB}_{\mathbf{k}}(\omega)$, $H^{BB}_{\mathbf{k}}(\omega)$, and $H^{AB}_{\mathbf{k}}(\omega)$  by interchanging the subscripts $A$ and $B$, respectively. The Green's function is $G_{\mathbf{k}}(\omega)=[G^{AA}_{\mathbf{k}}(\omega)+G^{AB}_{\mathbf{k}}(\omega)+G^{BA}_{\mathbf{k}}(\omega)+G^{BB}_{\mathbf{k}}(\omega)]/2$. Like the SF phase, the Green's function is decomposed as
\begin{eqnarray}
G_{\mathbf{k}}(\omega)=&\sum_{\alpha} \frac{C^p_{A,\alpha}(\mathbf{k})}{\omega-E^p_{A,\alpha}(\mathbf{k})}+\sum_{\alpha} \frac{C^p_{B,\alpha}(\mathbf{k})}{\omega-E^p_{B,\alpha}(\mathbf{k})} \nonumber \\
&-\sum_{\alpha} \frac{C^h_{A,\alpha}(\mathbf{k})}{\omega+E^h_{A,\alpha}(\mathbf{k})}-\sum_{\alpha} \frac{C^h_{B,\alpha}(\mathbf{k})}{\omega+E^h_{B,\alpha}(\mathbf{k})}.
\end{eqnarray}
The number of the poles of the Green's function  is $4 n_{\mathrm{max}}$ at zero temperature.
The momentum distribution is 
\begin{eqnarray}
n(\mathbf{k})=&\sum_{\alpha}C^h_{A,\alpha}(\mathbf{k})+\sum_{\alpha}C^h_{B,\alpha}(\mathbf{k}) \nonumber \\
&+N\left[ \left(\frac{\phi_A+\phi_B}{2}\right)^2 \delta_{\mathbf{k},\Gamma} +\left(\frac{\phi_A-\phi_B}{2}\right)^2 \delta_{\mathbf{k},\mathrm{M}} \right],
\end{eqnarray}
where $\mathrm{M}=(\pi, \pi)$ is the zone corner. The contribution of the condensate is derived in \ref{sec:con}. The ratio of the momentum distribution at $\mathrm{M}$ to that at $\Gamma$ is 
\begin{equation}
\frac{n(\mathbf{k}=\mathrm{M})}{n(\mathbf{k}=\Gamma)}=\left(\frac{\phi_A-\phi_B}{\phi_A+\phi_B}\right)^2 < 1.
\end{equation}
Thus, the peak at $\mathrm{M}$ is smaller than that at $\Gamma$.

\section{Excitation spectrum}
\label{sec:ex}

\begin{figure*}
\includegraphics{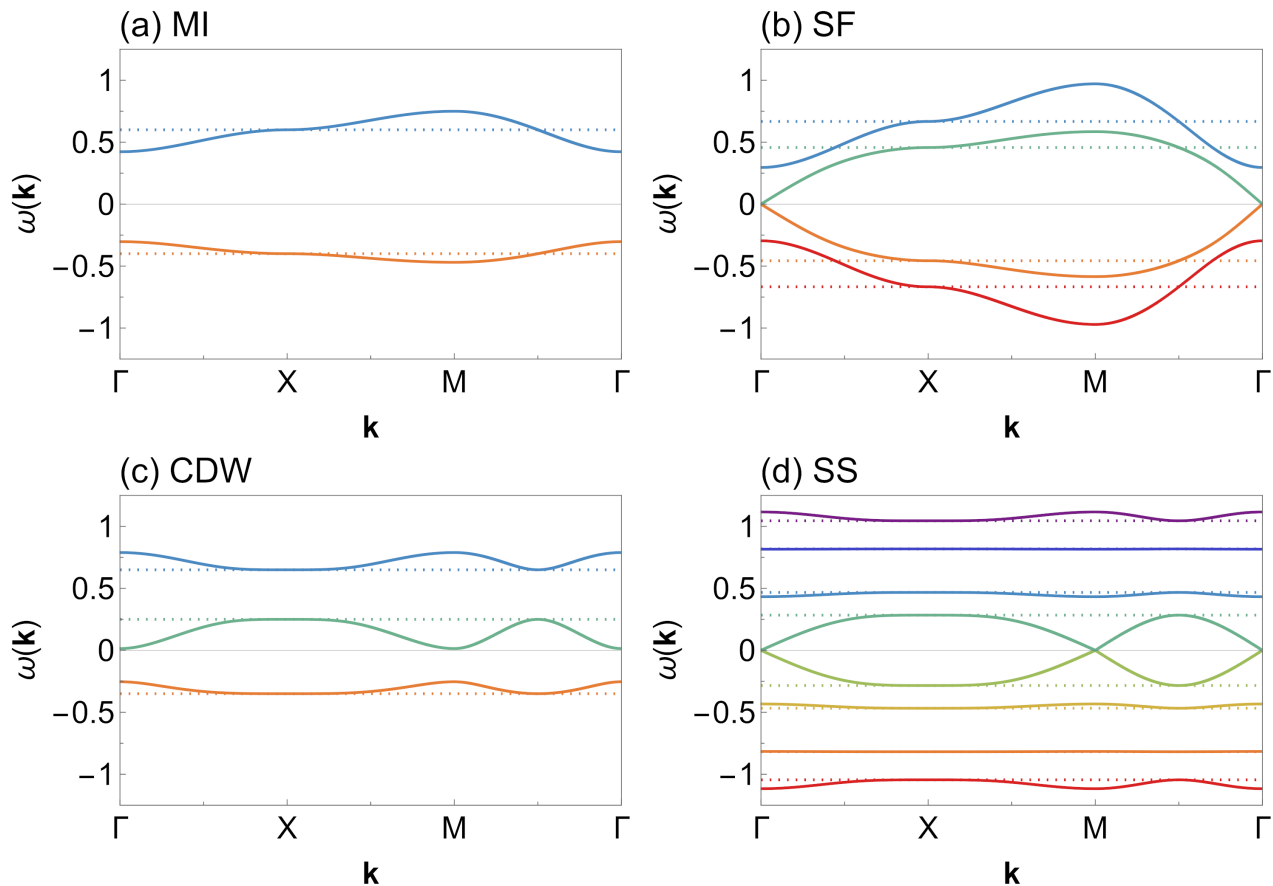}
\caption{\label{ex}Excitation spectrum of the ionic Bose-Hubbard model on a square lattice for $U=1$ at zero temperature for (a) MI ($\Delta=0$, $t/U=0.02$, $\mu/U=0.4$), (b) SF ($\Delta=0$, $t/U=0.05$, $\mu/U=0.4$), (c) CDW ($\Delta=0.3\,U$, $t/U=0.05$, $\mu/U=0.05$), and  (d) SS ($\Delta=0.3\,U$, $t/U=0.05$, $\mu/U=0.4$). The points $\Gamma$, X, and M are $(0,0)$, $(\pi,0)$, and $(\pi,\pi)$, respectively. The solid lines are the excitation energies calculated by the standard basis operator method, and the dotted lines are the mean-field excitation energies. Different colors correspond to different excitation modes.}
\end{figure*}

The excitation spectrum is given by the poles of the Green's function. \Fref{ex} shows the excitation spectrum of the ionic Bose-Hubbard model on a square lattice for $U=1$. Positive and negative excitation energies are particle and hole excitation energies, respectively.
The excitation energy calculated by the standard basis operator method agrees with the mean-field excitation energy at X, where the correction to the mean-field result is zero since $\epsilon_{\mathbf{k}}=0$ at X.  
 
In the MI phase, the spectrum is gapped for all $\mathbf{k}$. The particle excitation energy increases along $\Gamma$--X--M since $E^{p}(\mathbf{\mathbf{k}})$ is a monotonically increasing function of $\epsilon_{\mathbf{k}}$. The MI-SF phase transition occurs when either $E^{p}(\mathbf{k}=\Gamma)$ or $E^{h}(\mathbf{k}=\Gamma)$ becomes zero, which determines the MI-SF phase boundary.
 
In the SF phase, there are Goldstone modes, which are gapless and linear around $\Gamma$. The excitation energy of a gapless mode in the vicinity of $\Gamma$ is given by $E_g(\mathbf{k})=c\mathbf{k}$, where $c$ is the sound velocity. In addition to gapless modes, there exist gapped modes. There are a total of four modes since we diagonalize the mean-field Hamiltonian in the occupation number basis $\{|0\rangle, |1\rangle, |2\rangle \}$.
 
In the CDW phase, the hole excitation on sublattice $B$ is not possible since $n_B=0$. Thus, only the particle excitations on sublattices $A$ and $B$ and the hole excitation on sublattice $A$ are shown.  
The excitation energy is a function of $\epsilon_{\mathbf{k}}^2$, so the excitation spectrum along $\Gamma$--X is the same as that along M--X.

In the SS phase, since the excitation energy is a function of $\epsilon_{\mathbf{k}}^2$, the spectrum is gapless and linear around M as well as $\Gamma$. The gapless excitation at M reflects the simultaneous presence of superfluidity and solidity. In addition to a gapless particle excitation mode and a gapless hole excitation mode, there are three gapped particle excitation modes and three gapped hole excitation modes.

\section{Momentum distribution}
\label{sec:mo}

\begin{figure*}
\includegraphics{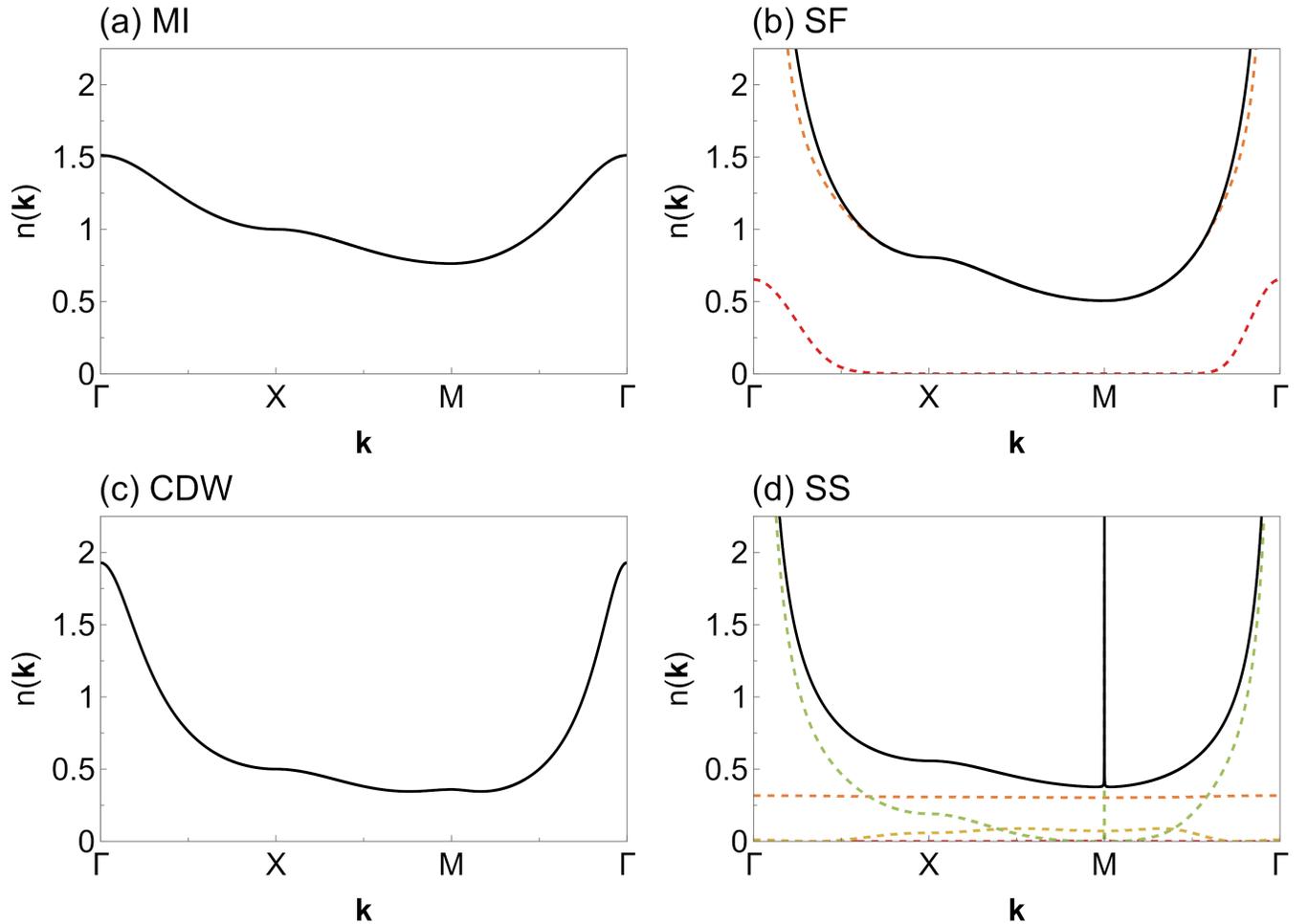}
\caption{\label{mo}Momentum distribution of the ionic Bose-Hubbard model on a square lattice for $U=1$ at zero temperature for (a) MI ($\Delta=0$, $t/U=0.02$, $\mu/U=0.4$), (b) SF ($\Delta=0$, $t/U=0.05$, $\mu/U=0.4$), (c) CDW ($\Delta=0.3\,U$, $t/U=0.05$, $\mu/U=0.05$), and  (d) SS ($\Delta=0.3\,U$, $t/U=0.05$, $\mu/U=0.4$).  The dashed lines are the spectral weights of hole excitation modes.}
\end{figure*}

The momentum distribution is given by the sum of the spectral weights of hole excitation modes. \Fref{mo} shows the momentum distribution of the ionic Bose-Hubbard model on a square lattice for $U=1$. In the MI phase, the momentum distribution decreases along $\Gamma$--X--M because $n(\mathbf{k})$ is a monotonically decreasing function of $\epsilon_{\mathbf{k}}$. 

In the SF phase, the momentum distribution is given by the sum of the spectral weights of the gapless hole excitation mode and that of the gapped hole excitation mode. The momentum distribution at $\mathbf{k}=\Gamma$ is given by the contribution of the condensate. It is proportional to the number of lattice sites $N$ and diverges in the thermodynamic limit ($N\rightarrow\infty$).  The momentum distribution increases drastically and diverges when $\mathbf{k} \rightarrow \Gamma$. This infrared divergence comes from the gapless excitation modes. When the gapless particle and hole excitation energies are given by $E^p_g(\mathbf{k})$ and $E^h_g(\mathbf{k})$, the Green's function is
\begin{equation}
G_{\mathbf{k}}(\omega)\propto\frac{1}{\omega-E^p_g(\mathbf{k})}\frac{1}{\omega+E^h_g(\mathbf{k})}.
\end{equation} 
Thus, the spectral weight of the gapless hole excitation mode is
\begin{equation}
C^h_g(\mathbf{k})=-\lim_{\omega\rightarrow -E^h_g(\mathbf{k})}[\omega+E^h_g(\mathbf{k})]G_{\mathbf{k}}(\omega) 
\propto\frac{1}{E^h_g(\mathbf{k})+E^p_g(\mathbf{k})}\propto\frac{1}{\mathbf{k}}.
\end{equation}
Therefore, the momentum distribution diverges when $\mathbf{k}\rightarrow\mathbf{0}$. This divergence is consistent with the analysis in the long wavelength limit~\cite{gavoret1964structure,griffin1993excitations}. The spectral weight of the gapped hole excitation mode also contributes to the momentum distribution around $\Gamma$, but that of the gapless hole excitation mode gives a major contribution.

In the CDW phase, in addition to a large peak at $\Gamma$, there is a small peak at M. This peak does not exist in the MI phase. The peak at M appears only close to the CDW-SS phase boundary. This behavior is similar to that of the extended Bose-Hubbard model~\cite{iskin2009momentum}. The critical value $t_m$ where $n(\mathbf{k})$ at M becomes a local maximum is the value which satisfies $\rmd n(\mathbf{k}=\mathrm{M})/\rmd t=0$.

In the SS phase, in addition to $\Gamma$, the momentum distribution diverges as $\mathbf{k}\rightarrow \mathrm{M}$ since the excitation is gapless at $\mathrm{M}$. The peak at $\mathrm{M}$ indicates the coexistence of superfluid and solid order. A peak at M is also observed in the quantum Monte Carlo simulations of the extended Bose-Hubbard model~\cite{ohgoe2012ground}.
The peak around $\mathrm{M}$ is narrower than that around $\Gamma$. This reflects that the peak at $\mathrm{M}$ is smaller than that at $\Gamma$. The spectral weight of the gapless hole excitation mode makes a great contribution to the momentum distribution around $\Gamma$ and $\mathrm{M}$. Away from $\Gamma$ and $\mathrm{M}$, the spectral weights of the gapped hole excitation modes give a major contribution. Thus, both gapless and gapped modes are important to understand the momentum distribution.

\section{Conclusions}
\label{sec:co}
By the standard operator method, we have derived Green's functions of the ionic Bose-Hubbard model in MI, SF, CDW, and SS phases. The excitation spectrum has gapped modes in all the phases and  gapless Goldstone modes in the SF and SS phases. We have shown that the momentum distribution has a peak at $\Gamma$ in all the phases and a peak at M in the SS phase and the CDW phase close to the CDW-SS phase boundary. Furthermore, we have demonstrated that the momentum distribution can be described by the excitation spectrum and spectral weights of hole excitation modes. We have clarified the contribution of both gapless and gapped hole excitation modes to the momentum distribution in the SF and SS phases.

The SBO method developed in this study will be applied to other lattices and contribute to understanding of Bose-Hubbard models.

\section*{Acknowledgments}
This work was supported by JST SPRING, Grant Number JPMJSP2108.

\appendix
\section{Contribution of the condensate}
\label{sec:con}
The Green's function for bosons at zero temperature is
\begin{equation}
G^{ij}(\omega)=\sum_{\alpha} \left(\frac{\langle g \vert b_i \vert \alpha \rangle \langle \alpha \vert b^{\dagger}_j \vert g\rangle }{\omega-E_{\alpha}}-\frac{ \langle \alpha \vert b_i \vert g \rangle \langle g \vert b^{\dagger}_j \vert \alpha \rangle }{\omega+E_{\alpha}}\right),
\end{equation}
where $\vert g \rangle$ is the ground state and $E_\alpha$ is the excitation energy of the eigenstate $\vert \alpha \rangle$. 
For $\alpha=g$, $E_\alpha=0$, and the term in the brackets vanishes. Thus, the momentum distribution of the condensate, which is the ground state in the SF and SS phases, can not be calculated from the Green's function. The contribution of the condensate is given by
\begin{equation}
n_c(\mathbf{k})=\frac{1}{N}\sum_{ij}\langle g \vert b^{\dagger}_i \vert g \rangle \langle g \vert b_j \vert g\rangle e^{i\mathbf{k}\cdot(\mathbf{R_{i}}-\mathbf{R_{j}})}.
\end{equation}
In the SF phase, the excitation energy is zero at $\mathbf{k}=(0,0)=\Gamma$, and $n_c(\mathbf{k}=\Gamma)=N\phi^2$, where $\phi=\langle g \vert b^{\dagger}_i \vert g \rangle = \langle g \vert b_i \vert g\rangle$.
In the SS phase, since sublattices A and B are inequivalent, 
\begin{eqnarray}
n_c(\mathbf{k})= \nonumber \\
\frac{1}{N}\left[ \sum_{i_Aj_A}\phi_A^2 e^{i\mathbf{k}\cdot(\mathbf{R_{i_A}}-\mathbf{R_{j_A}})} +\sum_{i_Bj_B}\phi_B^2 e^{i\mathbf{k}\cdot(\mathbf{R_{i_B}}-\mathbf{R_{j_B}})} \right. \nonumber \\
\left. +\sum_{i_Aj_B}\phi_A\phi_B e^{i\mathbf{k}\cdot(\mathbf{R_{i_A}}-\mathbf{R_{j_B}})} +\sum_{i_Bj_A}\phi_B\phi_A e^{i\mathbf{k}\cdot(\mathbf{R_{i_B}}-\mathbf{R_{j_A}})} \right],
\end{eqnarray}
where $\phi_m=\langle g \vert b^{\dagger}_{i_m} \vert g \rangle = \langle g \vert b_{i_m} \vert g\rangle$.
The excitation energy is zero at $\mathbf{k}=(0,0)=\Gamma$ and $\mathbf{k}=(\pi,\pi)=\mathrm{M}$.
For $\mathbf{k}=(0,0)$, $e^{i\mathbf{k}\cdot(\mathbf{R_{i_m}}-\mathbf{R_{j_n}})}=1$ for all sites, so 
\begin{equation}
n_c(\mathbf{k}=\Gamma)=N[(\phi_A+\phi_B)/2]^2.
\end{equation}
For $\mathbf{k}=(\pi, \pi)$, $e^{i\mathbf{k}\cdot(\mathbf{R_{i_m}}-\mathbf{R_{j_n}})}=1$ for $m=n$, and $e^{i\mathbf{k}\cdot(\mathbf{R_{i_m}}-\mathbf{R_{j_n}})}=-1$ for $m\neq n$, so 
\begin{equation}
n_c(\mathbf{k}=\mathrm{M})=N[(\phi_A-\phi_B)/2]^2.
\end{equation}
\section*{References}
\bibliographystyle{unsrt}
\bibliography{ibhm_sbo}

\end{document}